\documentclass{jps-cp}
\usepackage{txfonts} 
\title{$\Xi(1690)^-$ resonance production
via $K^-p \to  K^+K^-\Lambda$}

\author{Seung-il \textsc{Nam}$^{1,2}$ and Jung Keun \textsc{Ahn}$^{3}$}

\inst{$^{1}$Department of Physics and Institute for Radiation Science \& Technology (IRST), \\
Pukyong National University (PKNU), Busan 608-737, Republic of Korea \\
$^{2}$Asia Pacific Center for Theoretical Physics (APCTP), Pohang 790-784, 
Republic of Korea\\
$^{3}$Department of Physics, Korea University, Seoul 02841, Republic of Korea}

\email{sinam@pknu.ac.kr}

\recdate{January 18, 2019}

\abst{In this talk, we investigate $\Xi(1690)^-$ production from 
the $K^-p\to K^+K^-\Lambda$ reaction wit the effective 
Lagrangian method and consider the $s$- and $u$-channel $\Sigma/\Lambda$ ground states 
and resonances for the $\Xi$-pole contributions, 
in addition to the $s$-channel $\Lambda$, $u$-channel nucleon pole, 
and $t$-channel $K^-$-exchange for the $\phi$-pole contributions. 
The $\Xi$-pole includes $\Xi(1320)$, $\Xi(1535)$, $\Xi(1690)(J^p=1/2^-)$,
and $\Xi(1820)(J^p=3/2^-)$. We compute the Dalitz plot density of 
$(d^2\sigma/dM_{K^+K^-}dM_{K^-\Lambda}$ at 4.2 GeV$/c$) 
and the total cross sections for the $K^-p\to K^+K^-\Lambda$. 
Employing the parameters from the fit, we present 
the cross sections for the two-body 
$K^-p\to K^+\Xi(1690)^-$ reaction near the threshold. We also demonstrate that the Dalitz plot analysis for $p_{K^-}=1.915 \sim2.065$ GeV/c makes us to explore direct information for $\Xi(1690)^-$ production, 
which can be done by future $K^-$ beam experiments.}
\kword{Strangeness $S=-2$, $\Xi(1690)$, kaon beam, Dalitz process.}
\begin{document}
\maketitle
\section{Introduction}
$S=-2$ three-star baryon states include $\Xi(1690)^-$, 
$\Xi(1820)$($J^p=3/2^-$), $\Xi(1950)$, and $\Xi(2030)$. 
The third state of $\Xi$ has not yet been confirmed between
$\Xi(1620)$ and $\Xi(1690)^-$~\cite{ramos, garcia, oh, xiao, sekihara, miyahara, khem}. $\Xi(1690)^-$ is near the $\Sigma\overline{K}$ threshold, and its 
existence has been firmly established by several experiments 
\cite{dionisi, biagi, biagi2, belle}. Recently, the BaBar Collaboration \cite{babar} reported that $J=1/2$ assignment was favored for 
$\Xi(1690)^-$ from its decay angular distribution. 
The $\Xi(1690)^0$ was reconstructed from $\Lambda K_S^0$ 
in the $\Lambda_c^+\to \Lambda K_S^0 K^+$ decay. Nevertheless, its spin and parity 
have not yet been clearly determined. 

Experimentally, $\Lambda_c^+\to\Lambda K_S^0 K+$ is particularly
attractive, as high-statistics data are available from Belle/Belle-II 
and LHCb Collaborations. Nonetheless, the interference between $\Xi(1690)^-$
and $a_0(980)$ appears with a fixed crossing location in the phase space.
The phase in the interference between the two resonances could change the 
spin analysis result. Hence, it is necessary to carry out 
a $\Xi(1690)^-$ production experiment using 
the $(K^-,K^+)$ reaction near the threshold. $\Xi(1690)^-$ is 
produced in the $(K^-,K^+)$ reaction and decays to $\Lambda K^-$. 
In the $K^-p\to K^+K^-\Lambda$ reaction, the $\phi(1020)\to K^+K^-$ amplitude 
could interfere with the $\Xi(1690)^-$ production amplitude. 
However, the $\phi(1020)$ resonance is very narrow, so it can readily be
isolated from the $\Xi(1690)^-$ resonance. Moreover, 
the relative location of the interference region can change with
the $K^-$ beam momentum.  
   
In this talk, we provide numerical calculation results for the
production of $\Xi(1690)^-$ from the $K^-p\to K^+K^-\Lambda$ reaction within
the effective Lagrangian approach. We calculate the total and differential cross sections for 
the $K^-p\to \Xi(1690)^-K^+$ reaction 
in a beam momentum range from 2.1 GeV$/c$ to 2.3 GeV$/c$.
We also demonstrate that 
the Dalitz plot analysis of the $K^-p\to K^+K^-\Lambda$ reaction enables
us to access direct information concerning the $\Xi(1690)^-$ production. The details of the present talk can be found in Ref.~\cite{Ahn:2018hbc}.
\section{Theoretical framework}
\begin{figure}[!h]
\begin{center}
\includegraphics[width=10cm]{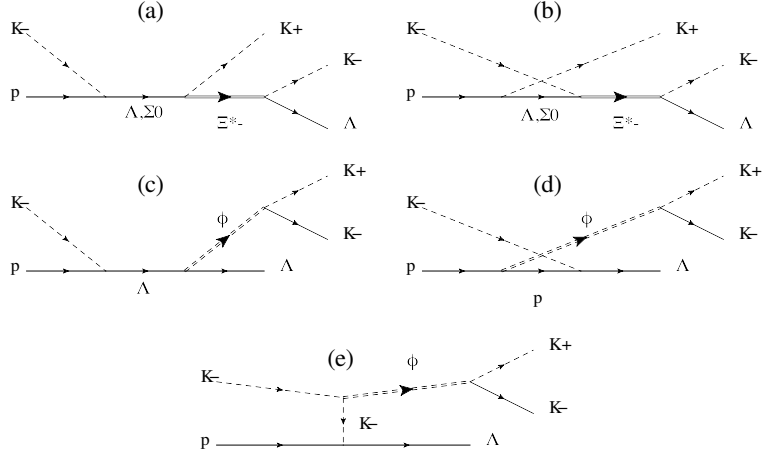}
\end{center}
\caption{Feynman diagrams for the $K^-p\to K^+K^-\Lambda$ reaction
at the tree-level Born approximation. 
Diagrams (a) and (b) contributed to the $\Xi$-pole, 
whereas (c), (d), and (e) contributed to the $\phi$-pole. 
The intermediate $\Lambda/\Sigma^0/\Xi^-$ denote the ground-states as well as the resonances.}       
\label{fig:diagram}
\end{figure}

For the $K^-p\to K^+K^-\Lambda$ reaction, the $s$- and $u$-channel diagrams 
are taken into account for the $\Xi$ production. Four $\Lambda$ states 
($\Lambda(1116)(J^p=1/2^+),~\Lambda(1405)(J^p=1/2^-),
~\Lambda(1520)(J^p=3/2^-)$, and~$\Lambda(1670)(J^p=1/2^-)$) and two
$\Sigma$ states ($\Sigma(1192)(J^p=1/2^+)$ 
and~$\Sigma(1385)(J^p=3/2^+)$) are included 
in the present calculation for $s$- and $u$-channel contributions.  
For the $\Xi$ production, four 
$\Xi$ states ($\Xi(1322)(J^p=1/2^+),~\Xi(1532)(J^p=3/2^+),
~\Xi(1690)(J^p=1/2^-)$, and $\Xi(1820)(J^p=3/2^-)$ are considered for the $\Lambda K^-$ decay channel. The relevant Feynman diagrams are depicted in Fig.~\ref{fig:diagram}.

Here, we assume that the $\Xi(1690)^-$ has a spin-parity of $J^p=1/2^-$, 
as suggested by theoretical works~\cite{xiao, khem} 
and reported by the BaBar Collaboration \cite{babar}. 
To compute the invariant amplitudes for 
the $K^-p\to K^+K^-\Lambda$ reaction, 
we use the effective Lagrangian densities for the interaction vertices 
as follows:
\begin{eqnarray}
\label{eq:EFFLAG}
\mathcal{L}_{KBB}&=&-ig_{MBB}(\overline{B}{\bf\Gamma})(\gamma_5K)({\bf\Gamma}B), \\
\mathcal{L}_{K\mathcal{B}B}&=&\frac{g_{K\mathcal{B}B}}{M_K}
(\overline{\mathcal{B}}_{\mu}{\bf\Gamma}\gamma_5)
(\gamma_5\partial^\mu K)({\bf\Gamma} B), \\
\mathcal{L}_{\phi KK}&=&-ig_{KK\phi}\phi^{\mu}
\left[(\partial_{\mu}K^{\dagger})K-(\partial_{\mu}K)K^\dagger\right]+\mathrm{h.c.},
\\
\mathcal{L}_{\phi BB}&=&-g_{\phi BB}\overline{B}
\left[\gamma^\mu- \frac{\kappa_{\phi BB}}{2M_B}
\sigma_{\mu\nu}\partial^\nu \right]\phi^*_\mu B+\mathrm{h.c.}, 
\end{eqnarray}
where $B$ and $\mathcal{B}$ stand for 
baryons with spin-$1/2$ and spin-$3/2$, respectively. 
We should mention that, in the present calculation, we did not consider 
the $K\mathcal{B}\mathcal{B}$ vertex for brevity, as there are no experimental data available for this reaction. 

The coupling constants for the ground-state hadron vertices, such as $g_{KN\Lambda(1116)}$, are taken 
from the prediction of the Nijmegen soft-core potential model (NSC97a)~\cite{Rijken:1998yy}. 
The coupling constants for the $s$-wave resonances, $\Lambda(1405)$ and $\Lambda(1670)$, 
are obtained from the chiral unitary model~\cite{Nakayama:2006ty}, where the resonances are generated dynamically 
by the coupled-channel method with the Weinberg--Tomozawa (WT) chiral interaction. 
The couplings for $\Xi(1690)$ and $\Xi(1820)$ are estimated by ChUM~\cite{khem} and 
the SU(6) relativistic quark model~\cite{xiao}, respectively. 

Regarding the coupling constants with two hyperon resonances, 
such as $g_{K\Lambda^*\Xi^*}$ and $g_{K\Sigma^*\Xi^*}$, 
there is no experimental nor theoretical information available. Furthermore, it is also difficult and uncertain 
to simply employ the flavor SU(3)-symmetry relation, which is used to obtain $g_{K\Lambda^*\Xi }$ and 
$g_{K\Sigma^*\Xi}$ as in Ref.~\cite{Shyam:2011ys}. Hence, we set those coupling constants to be zero for simplicity, 
although in practice their unknown contributions can be absorbed into the cutoff parameters of the form factors. 

We choose the phenomenological phase factors, $e^{3i\pi/2}$ and $e^{i\pi/2}$ for the amplitudes with
the spin-$1/2$ and spin-$3/2$ $\Xi$ hyperons, respectively, as follows:
\begin{eqnarray}
i\mathcal{M}_{\rm total}=ie^{i\pi/2}\mathcal{M}_{\Xi_{3/2}}+ie^{i3\pi/2}\mathcal{M}_{\Xi_{1/2}}+i\mathcal{M}_{\phi}.
\end{eqnarray} 
Note that these phase factors are determined to reproduce the experimental data~\cite{gay}. 

\section{Numerical results}
\begin{figure}[t]
\begin{tabular}{cc}
\includegraphics[width=8cm]{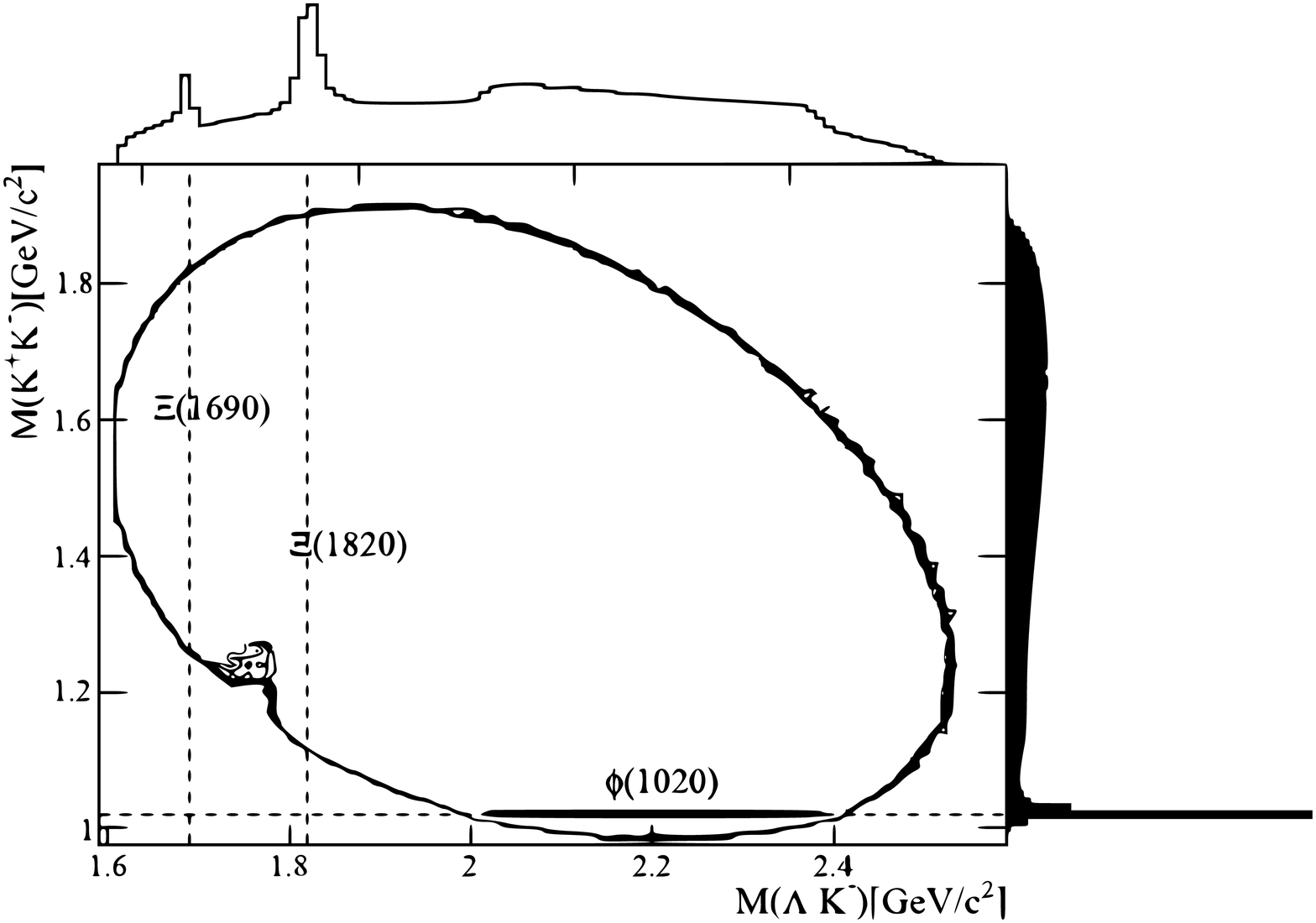}
\includegraphics[width=7cm]{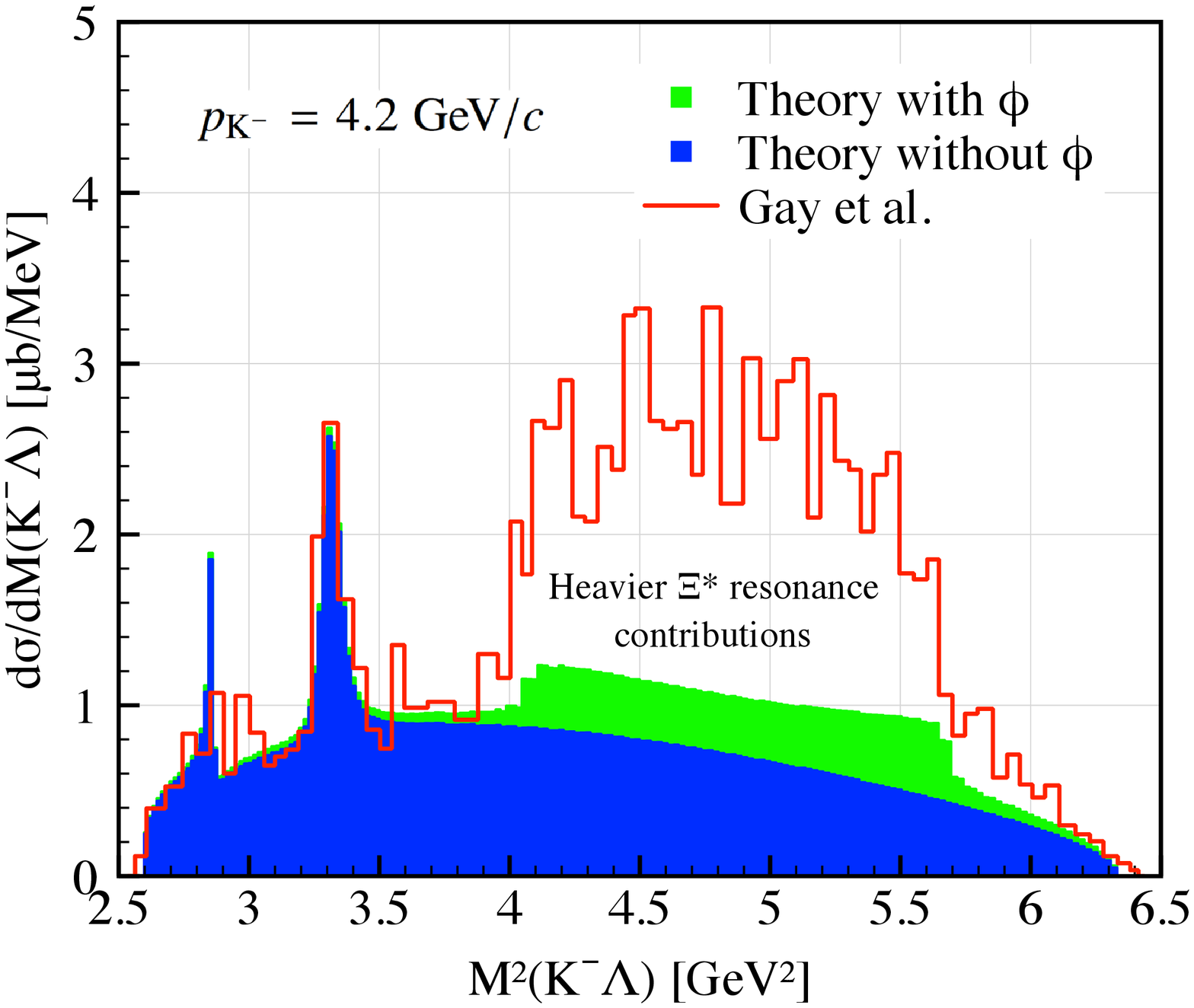}
\end{tabular}
\caption{(Color online) (Left) Calculated Dalitz plot density ($d^2\sigma/dM_{K^+K^-}dM_{K^-\Lambda}$) 
for the $K^-p\to K^+K^-\Lambda$ reaction at $p_{K^-}=4.2$ GeV. 
(Right) Differential cross section $d\sigma/dM_{K^-\Lambda}$ 
as a function of the invariant mass squared $M^2_{K^-\Lambda}$ 
at $p_{K^-}=4.2$ GeV. 
The green and blue areas indicate the results 
with and without the $\phi(1020)$ contribution, respectively. 
The experimental data~\cite{gay} are overlaid as a histogram.}       
\label{fig:gay}
\end{figure}

In this Section, we discuss the numerical results for the $\Xi(1690)$ production. We first show 
the numerical results for the $K^-p\to K^+K^-\Lambda$ reaction. 
The calculated Dalitz plot for the double differential cross section 
$d^2\sigma/dM_{K^+K^-}dM_{K^-\Lambda}$ at $p_{K^-}=4.2$ GeV$/c$ ($E_\mathrm{cm}=3.01$ GeV) 
is represented in the left panel of Fig.~\ref{fig:gay}, 
where the $\Xi^*(1690)$ and $\Xi(1820)$ resonances appear as vertical bands, while
$\phi(1020)$ appears as a horizontal band in the bottom side. At this energy, there is no interference effect 
between $\Xi^*$s and $\phi(1020)$. 
The Dalitz plot was projected on the $K^-\Lambda$ mass axis, as shown in the right panel of Fig.~\ref{fig:gay}. 
The experimental data are taken from Ref.~\cite{gay}, which is the only data set available so far 
for the $K^-p\to K^+K^-\Lambda$ reaction. The experiment was performed using the $K^-$ beam at 4.2 GeV$/c$ to study $\Xi(1820)$ and higher resonances. 
We then fit the data with the line shape of 
our calculation result in the low-mass region below $M^2_{K^-\Lambda}=3.3\,\mathrm{GeV}^2/c^4$. 

\begin{figure}[t]
\begin{tabular}{cc}
\includegraphics[width=7cm]{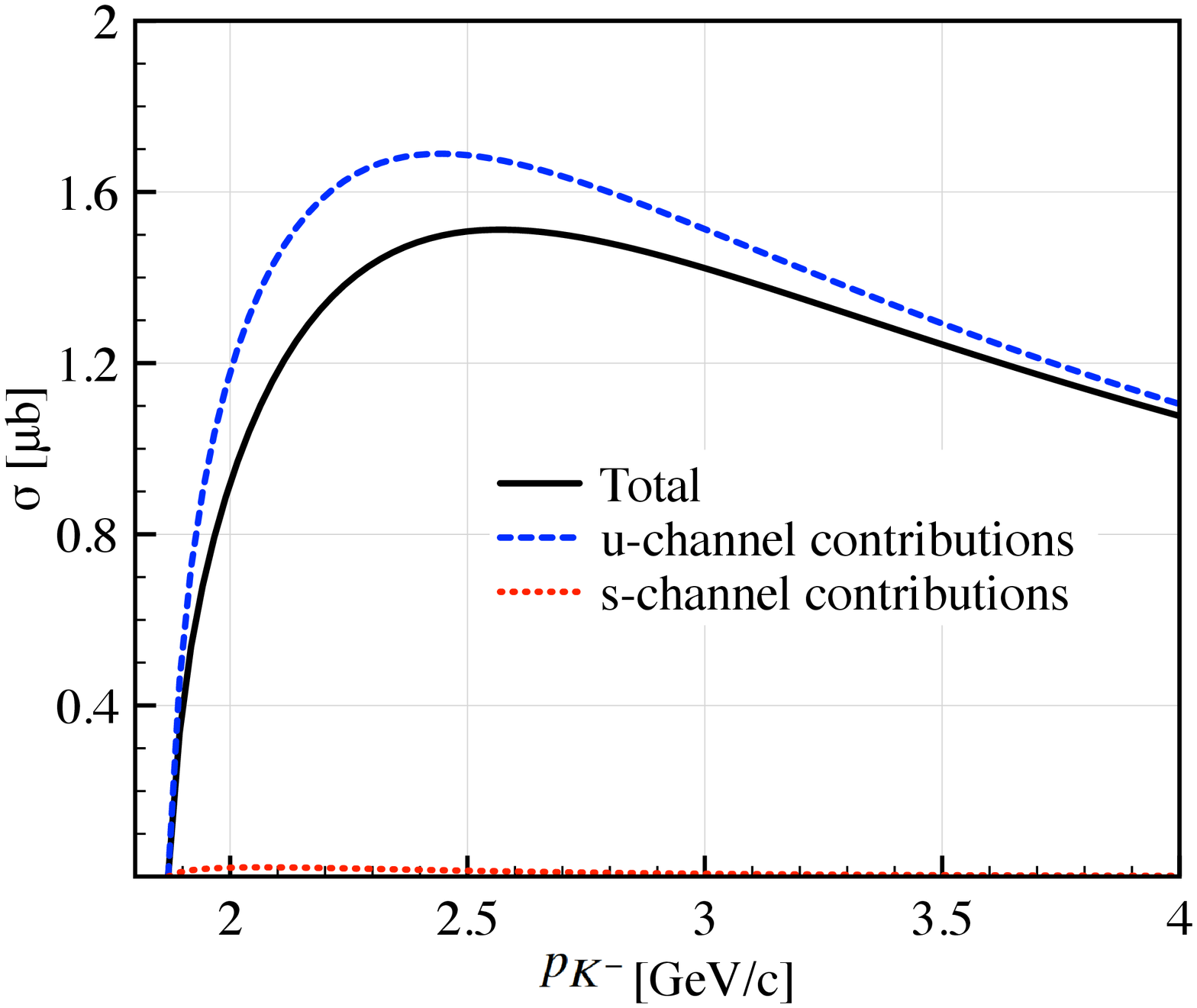}
\includegraphics[width=7cm]{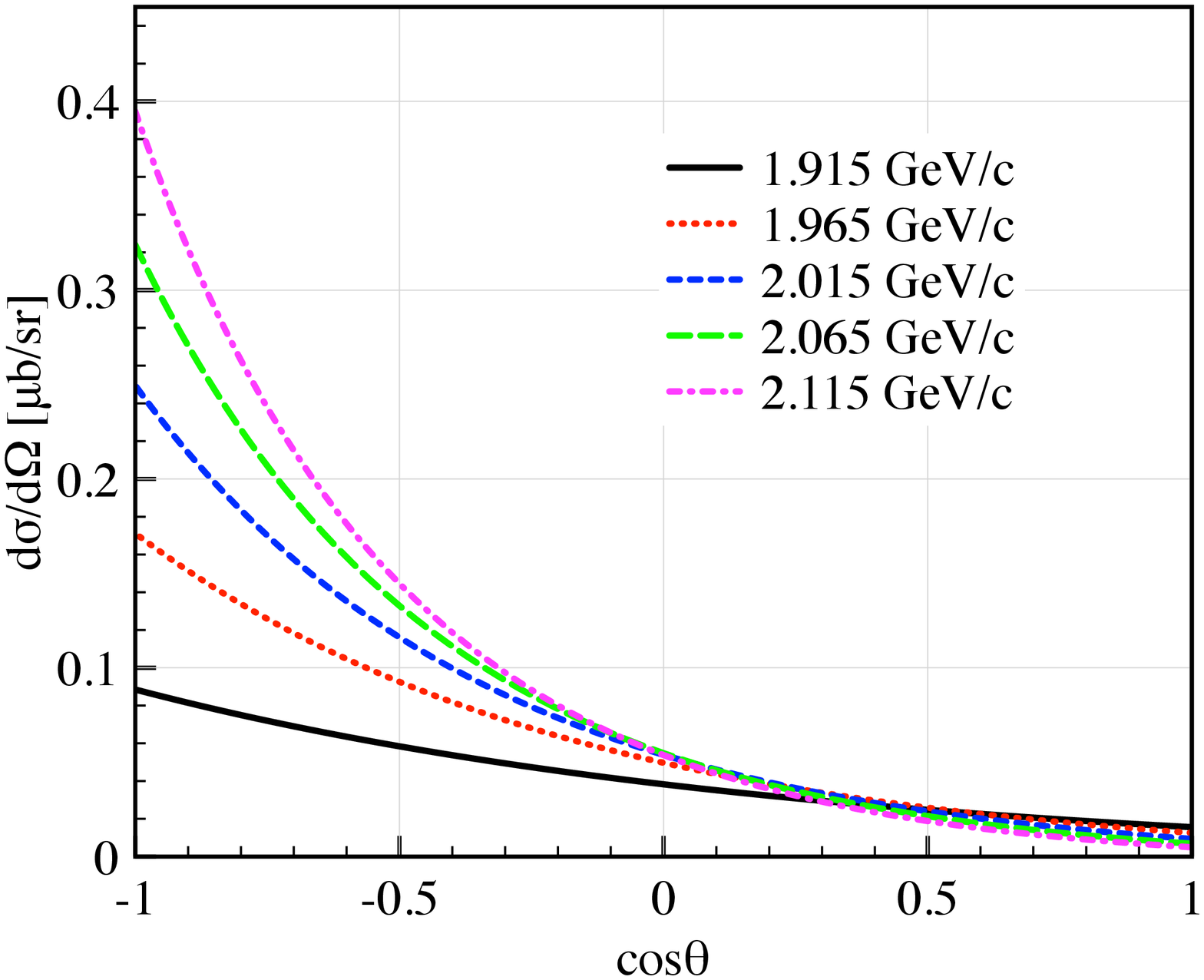}
\end{tabular}
\caption{(Color online) (Left) Total cross section for $K^-p\to K^+\Xi(1690)^-$ 
as functions of $p_{K^-}$. (Right) Differential cross section 
for $K^-p\to K^+\Xi(1690)^-$ as functions of the angle for the outgoing $K^+$ in the c.m. frame for several beam momenta $p_{K^-}$.}       
\label{fig:xross}
\end{figure}

After fixing the model parameters by fitting with the three-body experimental data, the total cross sections for $K^-p\to K^+\Xi(1690)^-$ are computed and represented as a function of $K^-$ beam momentum ($p_{K^-}$) from threshold
to 4 GeV$/c$ in the left panel of Fig.~\ref{fig:xross}. 
It increases rapidly from the threshold and peaks 
at $p_{K^-}=2.6$ GeV$/c$ ($E_{\rm cm}=2.47$ GeV) with $1.5~\mu$b, 
after which it decreases 
smoothly. As shown in the right panel of Fig.~\ref{fig:xross}, the $u$-channel contribution is much larger than 
the $s$-channel contribution. In our present calculation, 
we set the coupling constant $(g_{KY^*\Xi^*})$ to zero to avoid further theoretical uncertainty.   
Shyam {\it et al.} ~\cite{Shyam:2011ys} assumed that 
$g_{KY^*\Xi}=g_{KY^*N}$ for the $K^-p\to K^+\Xi^-$ reaction. 
However, there is no firmly established theoretical basis for 
the coupling constants $(g_{KY^*\Xi^*})$. 
\section{Summary}
In this talk, we present our recent work on the $\Xi(1690)^-$ production in the 
$K^-p\to K^+\Xi(1690)^-$ reaction within the effective Lagrangian approach. 
We consider the $s$- and $u$-channel $\Sigma/\Lambda$ ground states 
and resonances for the $\Xi$-pole contributions, 
in addition to the $s$-channel $\Lambda$, $u$-channel nucleon pole, 
and $t$-channel $K^-$-exchange for the $\phi$-pole contributions. 
The $\Xi$-pole includes $\Xi(1320)$, $\Xi(1535)$, $\Xi(1690)(J^p=1/2^-)$,
and $\Xi(1820)(J^p=3/2^-)$. We calculate the Dalitz plot density of 
$(d^2\sigma/dM_{K^+K^-}dM_{K^-\Lambda}$ at 4.2 GeV$/c$) and the total cross sections 
for the $K^-p\to K^+K^-\Lambda$ reaction near the threshold 
to determine the coupling constants and the form factors
for the two-body $K^-p\to K^+\Xi(1690)^-$ reaction.  
The calculated differential cross sections for the 
$K^-p\to K^+\Xi(1690)^-$ reaction near the threshold 
show a strong enhancement at backward $K^+$ angles, 
caused by the dominant $u$-channel contribution.

\end{document}